# DIFFUSION PATHS BETWEEN PRODUCT LIFE-CYCLES IN EUROPEAN PHONOGRAPHIC MARKET

**Andrzej Buda#, Andrzej Jarynowski#***

\# Institute of Interdisciplinary Research, Wrocław, Poland
\* Smoluchowski Institute of Physics, Jagiellonian University, Kraków, Poland

We have investigated the product life-cycles of almost 17 000 hit singles performed on the 12 biggest national phonographic markets in Europe including: Austria, Belgium, France, Germany, Ireland, Italy, Netherlands, Norway, Spain, Sweden, Switzerland and the United Kingdom. We have considered weekly singles charts from the last 50 years (1966-2015) in each country. We analyze the spread of hit singles popularity (chart topping) as an epidemiological process performed on various European countries. Popular hit singles are contagious from one country to another. Thus, we consider time delays between the initial hit single release and reaching the highest position on consecutive national singles charts. We create directed network of countries representing transitions of hit singles popularity between countries. It is obtained by simulating the most likely paths and picking up the most frequent links. A country of initial hit single release is considered as a source of infection. Our algorithm builds up spanning trees by attaching new nodes. The probability of attachment depends on:
    1) new node's immunity
    2) infectivity of previous nodes from the tree.
Thus we obtain network of popularity spread with: a hub – the UK, a bridge – the Netherlands and outliers – Italy and Spain. We have found a characteristic topology of hit singles popularity spread. The positive correlation between this network and geographic or cultural grid-map of Europe is also observed. However, the network of popularity spread has some typical properties of complex networks.

1. Introduction

      Statistical properties of the global phonographic market has been already investigated well by methods of econophysics where the value of an artist has been defined by weekly albums sales. This system is more predictable than financial markets and more complex because of record labels, artists, mass media and additional seasonal groups of collective customers that buy long-playing records. Thus, product life-cycles expressed by trajectories of weekly albums sales reach the maximum in the first week of record sales after the premiere and decrease affected by random processes (Jarynowski and Buda, 2014).
    However, there is another part of phonographic market represented by songs released as a hit singles. The spread of hit singles popularity over countries reminds classical product life-cycles because the official definition of popularity in case of songs is more complex. National charts have been usually based on weekly physical singles sales, but smaller markets have also compiled official national charts according to official airplays on the radio or TV. On the Internet era (2003 until now), almost all national hit singles charts are mostly based on digital streaming of

songs (mp3, ringtones, etc.) Popularity of songs as a compilation of these factors is complex, so in our research we decided to measure this phenomena from economic point of view (value of a song is measured weekly by positions on national charts). We have investigated trajectories of almost 17 000 hit singles performed on the 12 biggest national phonographic markets in Europe including: Austria (A), Belgium (B), France (F), Germany (D), Ireland (IRE), Italy (I), Netherlands (NL), Norway (N), Spain (E), Sweden (S), Switzerland (CH) and the United Kingdom (GB). We have considered weekly singles charts from the last 50 years (1966-2015) in each country. The process of information spread have been already investigated and described from sociological and computational science's point of view where European music charts network, positive correlation with geographical and cultural distance network has been explored in various ages (Buda and Jarynowski, 2015).

In this paper, we continue our research and focus on economic properties and interactions between national markets. Classic product life-cycle theory is defined by stages that depend on diffusion of innovations (including: innovators, early adopters, early majority, late majority and laggards) (Rogers, 1962). In case of hit-singles, the national charts are uncomparable because of complexity (different methods of compilations in each country, size of a market, etc.). But if we consider stages of popularity for a single performed on various countries (that represent innovators, early adopters and late adopters), the product life-cycle may be described epidemiologically by SI models (Anderson and May, 1992). Our inspiration also comes from geographical dependencies between currencies where the rate returns usually affects other currencies in the neighbourhood because of triangular arbitrage and the existence of the Epp's effect on much longer time scales (Rogers, 1962). Like in the foreign exchange markets, popular hit singles are contagious from one country to another. Thus, we describe stages of product life-cycle in terms of epidemiology and consider time delays between the initial hit single release and reaching the highest position on consecutive national singles charts. Our aim is to show paths of interactions between the 12 biggest national markets in Europe and detect how one country affect another.

2. Data analysis

At the very beginning, it is necessary to define the state of popularity for a song as the highest position in a national singles chart. For example, 'We Found Love' performed by Barbadian singer Rihanna in 2011 initially entered the singles charts on October the $8^{th}$.

Table 1. Higher and higher chart positions in the national charts by Rihanna's 'We Found Love' within 8 weeks. After reaching the top position, cells are matched in black because a country is infected.

| Country | 1$^{st}$ week | 2$^{nd}$ week | 3$^{rd}$ week | 4$^{th}$ week | 5$^{th}$ week | 6$^{th}$ week | 7$^{th}$ week | 8$^{th}$ week |
|---------|---------|---------|---------|---------|---------|---------|---------|---------|
| A | - | - | - | 12 | 12 | 4 | ■ | ■ |

| B   | 3  |    |    |   |   |   |   |   |
|-----|----|----|----|---|---|---|---|---|
| CH  | 3  | 3  | 2  | 2 | 2 | 1 |   |   |
| D   | -  | -  | -  | 1 |   |   |   |   |
| E   | 15 | 14 | 14 | 9 | 9 | 9 | 5 | 3 |
| F   | 1  |    |    |   |   |   |   |   |
| GB  | -  | 1  |    |   |   |   |   |   |
| I   | -  | -  | -  | 7 | 4 | 4 |   |   |
| IRE | -  | 3  | 1  |   |   |   |   |   |
| N   | 1  |    |    |   |   |   |   |   |
| NL  | 14 | 3  |    |   |   |   |   |   |
| S   | 2  | 2  | 2  | 2 | 1 |   |   |   |

In this case, the source of infection was Belgium, but in the first week France and Norway had been already infected. Italy and Spain were the last countries that fulfilled this spread of global popularity. If we consider the time delay between the first week of entering the charts and the week of reaching the highest positions, we will obtain the map of Europe (Fig. 1) according to investigated data set (1966-2015) of hit singles. Statistically, the UK, the Netherlands and Belgium usually start the infection.

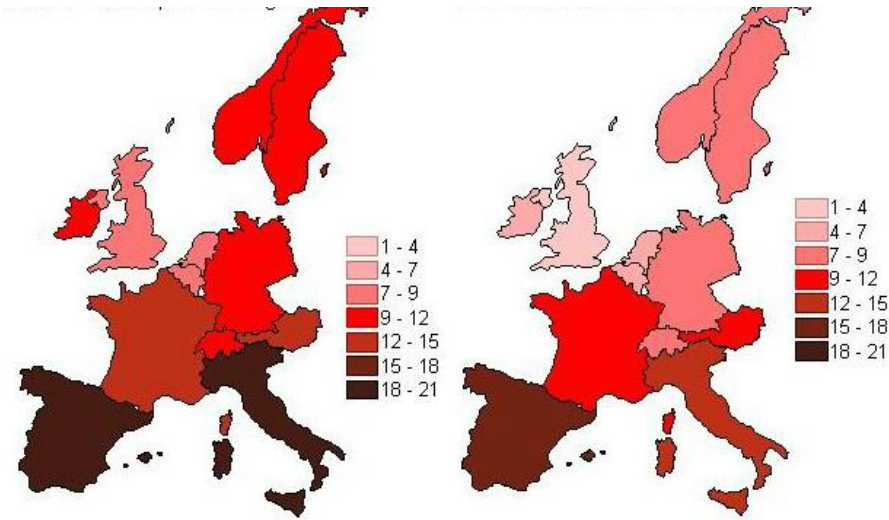

Figure 1. Mean (left) and median (right) time of achieving the highest positions on the charts (Buda and Jarynowski, 2015).

For each song, it is possible to detect the Minimum Spanning Tree - MST (Fig. 2) according to the matrix of individual time delays defined by distances:

$$d_{xy} = | t_x - t_y | \qquad (1)$$

where $t_x$ and $t_y$ represents weeks of reaching the highest positions in countries X and Y.

Deterministic construction of a chain of events like MST does not reveal unique solution. It is degenerated, because in the same time more than one country could adopt new single (see ABBA's case on Fig. 2). The MST also works as a Markovian process, and the history is going to be forgotten. We provide an algorithm to retrieve set of allowed trees, where an actual connection depends on the whole history.

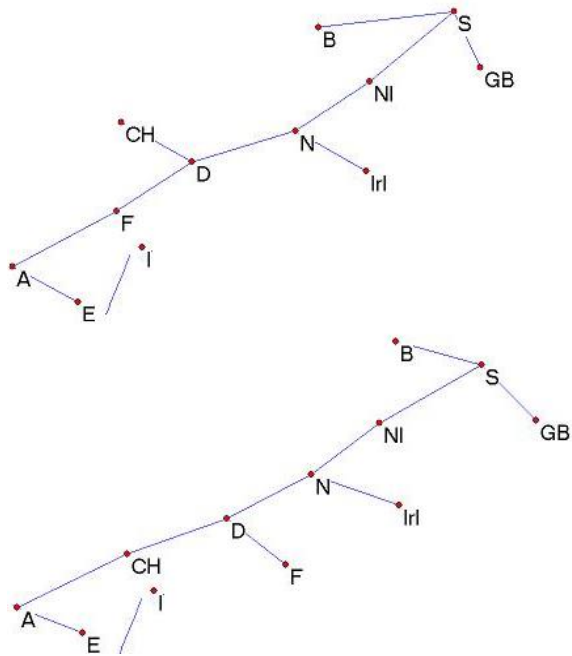

Fig. 2 The Minimum Spanning Tree based on matrix of time-delays between countries for ABBA's single 'Dancing Queen' (1976). The source of infection was in Sweden.

2. Results

In our research, we have detected the sources of infections (initial countries) for all the 50 most popular hit singles that finally affected the whole Europe (1966-2015). Then we have created the most likely paths of infection. Our algorithm builds up spanning trees for each hit-single 1000 times by attaching new nodes:

1) We choose source of infection - country of initial release (e.g. ABBA's song 'Dancing Queen' comes from Sweden) [Fig. 2];

2) We attach next in row node to the source (e.g. Belgium and the UK simultaneously are attached to Sweden) [Fig. 2];
3) The next ones (X) are added to one of the previous nodes (respectively to the size of a market in a previous node). The probability of attachment is inversely proportional to difference between time delays according to formula:

$$P_{x-y} \sim \frac{1}{t_x - t_y} * \frac{N_Y}{N_{AVG}} \qquad (2)$$

where:
$P_{x-y}$ - probability that new node X will be attached to node Y
$N_Y$ - market size of node Y
$N_{AVR}$ - average market size in Europe

In Western European countries, a phonographic market size $N_Y$ may be taken as a population Y size [10]. Probability of attachment increases in inverse proportion to time delays (memory effect). We weight each link in inverse proportion on steps of simulation to valuate fashion adoption in early steps (early adopters).

Table. 2 Node's degree calculated via simulation of popularity spread

| degree | out | in |
|--------|-----|-----|
| A      | 23  | 36  |
| B      | 60  | 65  |
| CH     | 45  | 63  |
| D      | 75  | 53  |
| E      | 16  | 27  |
| F      | 43  | 50  |
| GB     | 192 | 36  |
| I      | 17  | 36  |
| Irl    | 43  | 89  |
| N      | 49  | 65  |
| Nl     | 74  | 94  |
| S      | 55  | 78  |

Thus, we obtain a directed network of popularity spread with: a hub – the UK, a bridge – the Netherlands and outliers – Italy and Spain (Fig. 3). We have found a characteristic topology of hit singles popularity spread. It is clearly visible from which country viruses are able to be contagious to another. There is no surprise that the UK has the strongest ability to create (but not receive) viruses (due to highest out degree – Tab.2) from all over the European world. The Netherlands are most likely to adopt (be infected) be new fashion (due to highest in degree – Tab.2).

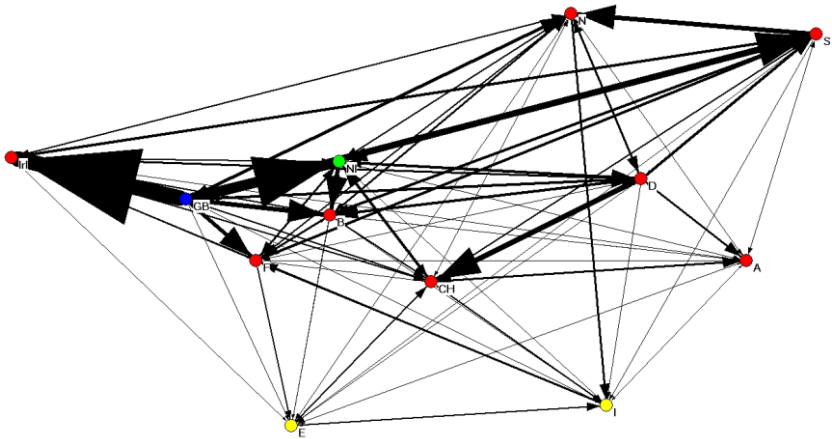

Fig. 3 The directed European network of popularity spread based on matrix of time-delays between countries. Network of popularity spread with: a hub – the UK, a bridge – the Netherlands and outliers – Italy and Spain.

The trees based on matrix of time delays reflect nonuniqueness. It occurs even for deterministic process of attachment. Clusterization and community detection (modularity) can bring two subnetworks:
–       Scandinavia, Benelux, UK, Ireland and Germany
–       Spain, Italy, Switzerland and Austria

The first one contains countries that easy catch all infections, the second one is more conservative. However, France belongs once to infectious community once to resistance community. The position of France in directed networks (Fig. 4) depends on the algorithm.

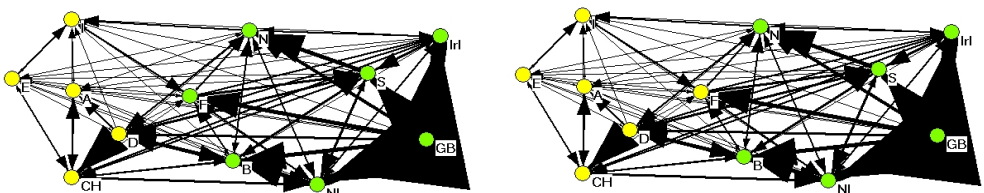

Fig. 4 The directed France misclassified - once in resistance community once in infectious community. Left: Louvain algorithm. Right: VOS Clustering.

4. Conclusions

In this paper we have shown the structural dependencies between various local European phonographic markets. We have built epidemiological simulation to extract the most important viral component of the system. We have received network of popularity spread with: a hub – the UK, a bridge – the Netherlands and outliers – Italy and Spain. The most influential country is the United Kingdom that infects others in early steps of propagation due to the highest out-degree [Tab.2]. However, the UK is immune to foreign hit-singles, especially from non English speaking countries. This node has one of the lowest in-degree in the whole Europe [Tab.2]. On the other hand, The Netherlands avoided to be the initial source of infection and could adopt foreign hit-singles quickly, no matter where they come from. The Dutch phonographic market is similar to European bridge (broker) (Przybyła and Weron, 2014) because of its high in-degree. These results reveal interactions between local phonographic markets in Europe because the process of popularity spread has some typical properties of complex networks (evidence of a hub). Diffusion and spreading processes has been investigated on empirical and random networks (Leskovec et all, 2007), (Boss et all, 2004).

We observe that the network of countries breaks down into 2 well-defined clusters [Fig. 4] of early and late adopters like other systems which has associated a meaningful taxonomy (Paulus and Kristoufek, 2015). However, the role of France in European market is still worth to explain. According to our results, France belongs once to infectious community, once to resistance community. This behaviour may be a result of broadcasting regulations in France where the spread of popularity through radio or TV for non-French songs is strictly limited to 50%. These limits have influence on hit-singles sales too. Thus, the French phonographic market is resistant, but it is able to adopt quickly the most popular foreign songs. In our research, however, we are limited to direct paths of interactions without considering any external fields.

.